\documentclass[epj]{svjour}
\usepackage{graphics}
\usepackage{graphicx}
\usepackage{amsmath}
\usepackage{color}
\usepackage{cite}
\usepackage{microtype}
\usepackage{hyperref}
\hypersetup{colorlinks=true,linkcolor=blue,citecolor=blue,menucolor=black,urlcolor=magenta,filecolor=blue}

\graphicspath{{figures/}}

\newcommand{\GeV}{\,\mbox{GeV}}

\newcommand{\blue}[1]{{\color{blue}#1}}

\newcommand{\CASA}{{Center for Advanced Simulation and Analytics (CASA), Forschungszentrum Jülich, 54245 Jülich Germany}}
\newcommand{\itp}{{CAS Key Laboratory of Theoretical Physics,
            Institute of Theoretical Physics,\\ Chinese Academy of Sciences,
            Beijing 100190, China}}
\newcommand{\bonn}{{Helmholtz-Institut f\"ur Strahlen- und
             Kernphysik and Bethe Center for Theoretical Physics,\\
             Universit\"at Bonn,  D-53115 Bonn, Germany}}
 
 \newcommand{\JSC}{{J\"{u}lich Supercomputer Centre, 
           Forschungszentrum J\"ulich, D-52425 J\"ulich, Germany}}
\newcommand{\ias}{{Institute for
           Advanced Simulation, 
           Forschungszentrum J\"ulich, D-52425 J\"ulich, Germany}}
\newcommand{\ucas}{{School of Physical Sciences,
            University of Chinese Academy of Sciences,
            Beijing 100049, China}} 
\newcommand{\peng}{{Peng Huanwu Collaborative 
Center for Research and Education, Beihang University, Beijing 
100191, China}}

\begin{document}
\title{Exclusion of a diquark--anti-diquark structure for the lightest positive-parity charmed mesons}

\author{Eric B.~Gregory\inst{1,2} \and Feng-Kun~Guo\inst{3,4,5} \and Christoph~Hanhart\inst{6,7} \and Stefan Krieg\inst{1, 2, 7} \and Thomas Luu\inst{6,7}
}

%
%
\institute{\JSC  \and \CASA \and \itp \and \ucas \and \peng \and \ias \and \bonn}
 %
%
\abstract{
The nature of low-lying scalar and axial-vector charmed mesons has been debated for decades, with hadronic molecular and compact tetraquark models being prominent candidates.
These two models predict quite different features  for the  accessible SU(3) multiplets in the scalar and axial-vector sectors, which can be tested through lattice calculations at SU(3) symmetric points.
In this work, we perform lattice calculations for both scalar and axial-vector charmed mesons with an SU(3) symmetric pion mass about 613~MeV for the SU(3) $[6]$ and $[\overline{15}]$ multiplets. We find that the $[6]$ multiplet exhibits attractive interactions in both scalar and axial-vector sectors, while the $[\overline{15}]$ multiplet shows repulsive interactions in both sectors. The energy shifts in the scalar and axial-vector sectors are compatible with each other within uncertainties.
These results are fully consistent with the hadronic molecular picture, while challenging the compact tetraquark model, which predicts the existence of low-lying $[\overline{15}]$ states in the axial-vector sector but not in the scalar sector.
}
\PACS{
      {12.38.Gc}{Lattice QCD calculations}  \and
      {13.75.Lb}{Meson-meson interactions}  \and
      {14.40.Lb}{Charmed mesons}
     } 

\maketitle

\section{Introduction}
\label{intro}
Because of color confinement of quantum chromodynamics (QCD), understanding the spectrum of hadrons is one of the most challenging tasks in the study of the strong interaction. Since 2003, when the $B$-factories entered the hunt for
the hadronic spectrum,
many hadrons were observed with properties in conflict with the predictions from conventional quark models that identify mesons as $\bar qq$ states. 
For example, the $D_{s0}^*(2317)$~\cite{Aubert:2003fg} and $D_{s1}(2460)$~\cite{Besson:2003cp} are significantly
lighter than the ground state scalar and axial-vector $c\bar s$ states at $2.48\GeV$ and $2.55\GeV$, respectively, predicted in the Godfrey-Isgur quark model~\cite{Godfrey:1985xj}.
This led to the development of various models, including $D^{(\ast)}K$ hadronic molecules~\cite{Barnes:2003dj,vanBeveren:2003kd,Szczepaniak:2003vy,Kolomeitsev:2003ac,Chen:2004dy,Guo:2006fu,Guo:2006rp,Gamermann:2006nm}, compact tetraquark states~\cite{Cheng:2003kg,Maiani:2004vq}, and mixtures of $c\bar{q}$ with tetraquarks~\cite{Browder:2003fk,Dai:2006uz}. 

The corresponding charm-nonstrange states are known as $D_0^*(2300)$~\cite{Abe:2003zm,Link:2003bd}
and $D_1(2430)$~\cite{Abe:2003zm}.
Clearly their masses are significantly higher than expected from the SU(3) breaking pattern, since they are too close to those of the partner states
containing strangeness.
This puzzle and the fact that the masses
of the $D_{s0}^*(2317)$ and $D_{s1}(2460)$ are equidistant
to the $DK$ and $D^*K$ thresholds, respectively,
are naturally understood when employing unitarized chiral perturbation theory (UChPT). 
For the singly heavy states this approach was pioneered in Ref.~\cite{Kolomeitsev:2003ac}---for more recent developments see, e.g., Refs.~\cite{Albaladejo:2016lbb,Du:2017zvv,Guo:2017jvc}. This formalism
allows one to calculate the nonperturbative dynamics of the scattering of the lightest pseudoscalar meson octet, the pseudo-Nambu-Goldstone bosons of the spontaneously broken chiral symmetry with the $D_{(s)}^{(*)}$ mesons in a controlled way. 
In particular, the coupling of SU(3) flavor anti-triplet ($D^{(*)+}$, $D^{(*)0}$ and $D_s^{(*)+}$) with the light pseudoscalar meson octet forms three irreducible representations (irreps)~\cite{Kolomeitsev:2003ac,Hyodo:2006kg,Albaladejo:2016lbb}: $$[\overline 3]\otimes [8]=[\overline 3]\oplus[6]\oplus[\overline{15}] \ .$$
The leading order chiral interaction from the Weinberg-Tomozawa term predicts that the interaction in the $[\overline 3]$ is attractive, in the $[6]$ is less attractive and the one in the $[\overline{15}]$ is repulsive. It is a necessity of the consistency of the formalism that this feature persists also when higher orders
are included---a feature confirmed in the corresponding calculations.
It is worth noting that only the $[\overline{3}]$ irrep exists for conventional $c\bar q$ mesons, and the presence of $[6]$ and $[\overline{15}]$ requires at least an additional $q\bar q$ pair, which can occur in both hadronic molecular and compact tetraquark models.

In Ref.~\cite{Liu:2012zya} a fit to lattice data for various light pseudoscalar meson--$D_{(s)}$-meson
scattering lengths (but not in the channel where
the $D_{s0}^*(2317)$ is located) fixed the a priori unknown low-energy constants in the next-to-leading order chiral Lagrangian. The resulting amplitudes
not only reproduced the correct $D_{s0}^*(2317)$ mass,
but also predicted its pion mass dependence~\cite{Du:2017ttu} that agrees well with lattice results~\cite{Bali:2017pdv}. 
In addition, this study solved the mass hierarchy puzzle mentioned above by confirming that the structure known as
$D_0^*(2300)$ in the Review of Particle Physics (RPP)~\cite{ParticleDataGroup:2024cfk} emerges from two distinct poles, one lighter and one heavier, in line with other findings~\cite{Kolomeitsev:2003ac,Guo:2006fu,Guo:2009ct,Guo:2015dha,Albaladejo:2016lbb,Guo:2018tjx,Guo:2018kno,Guoa:2018dhm}. 
Their pole locations are found at 
$\left(2105^{+6}_{-8}{-}i\,102^{+10}_{-11}\right)\text{MeV}$ and $\left(2451^{+35}_{-26}{-}i\, 134^{+7}_{-8}\right)\text{MeV}$~\cite{Albaladejo:2016lbb,Du:2017zvv}, respectively. The SU(3) partner of the $D_{s0}^*(2317)$ is the lighter one, which restores the expected mass hierarchy---they form a complete SU(3) anti-triplet. 
The heavier pole on the other hand is a member of SU(3) sextet, which is exotic as mentioned above.
Experimental support for the presence of two poles comes from an analysis of the high-quality LHCb data on the decays $B^-\to D^+\pi^-\pi^-$~\cite{Aaij:2016fma}, $B_s^0\to\bar{D}^0K^-\pi^+$~\cite{Aaij:2014baa},
$B^0\to \bar{D}^0 \pi^-\pi^+$~\cite{Aaij:2015sqa}, $B^-\to D^+\pi^- K^-$~\cite{Aaij:2015vea}, and
$B^0\to\bar{D}^0\pi^- K^+$~\cite{Aaij:2015kqa} performed
in Refs.~\cite{Du:2017zvv,Du:2019oki,Du:2020pui}, as well as from the fact that their existence is consistent with the lattice energy levels~\cite{Liu:2012zya,Mohler:2013rwa,Lang:2014yfa,Moir:2016srx,Bali:2017pdv,Cheung:2020mql} for the relevant two-body scattering~\cite{Albaladejo:2016lbb,Albaladejo:2018mhb,Guo:2018tjx}. Furthermore, the prediction that the $D\bar K$ is a  virtual state~\cite{Albaladejo:2016lbb} 
in the sextet and that the lightest $D_0^*$ should be significantly lighter than 2.3~GeV were also confirmed in lattice QCD (LQCD)~\cite{Moir:2016srx,Gayer:2021xzv},\footnote{The pole position in the LQCD analysis of Ref.~\cite{Mohler:2013rwa} is at 2.12(3)~GeV~\cite{Bulava:2022ovd}, consistent with predictions in Refs.~\cite{Albaladejo:2016lbb,Du:2017zvv}.} and the predicted lowest-lying bottom-strange scalar and axial-vector mesons~\cite{Guo:2006fu,Guo:2006rp} (for an update, see~\cite{Fu:2021wde}), as heavy quark flavor symmetry partners of the $D_{s0}^*(2317)$ and $D_{s1}(2460)$ states, agree well with lattice QCD results~\cite{Lang:2015hza}. 
In addition, it is shown in Ref.~\cite{Du:2020pui} that the $D_0^*(2300)$ with resonance parameters listed in the RPP~\cite{Zyla:2020zbs} is in conflict with the LHCb data on $B^- \to D^+\pi^-\pi^-$~\cite{Aaij:2016fma}, contrary to the two $D_0^*$ states scenario.
This two-pole structure indeed emerges as a more general pattern in the hadron spectrum, see, e.g., Ref.~\cite{Meissner:2020khl}.

One more piece of evidence for the existence of the sextet structure comes from the recent LHCb observation of a signal for an iso-vector exotic resonance with a mass around 2327~MeV~\cite{LHCb:2024iuo}. Such an iso-vector pole was first predicted in Ref.~\cite{Guo:2009ct} to be located on a remote Riemann sheet,\footnote{The statement in Ref.~\cite{Albaladejo:2016lbb} ``we do not find any pole that can be associated to a physical state'' was because the pole was located deep on a remote Riemann sheet.} having a real part around 2.3~GeV and a sizable imaginary part.
It is in the SU(3) sextet within the SU(3) symmetric limit and receives a mixture from the $[\overline{15}]$ representation due to SU(3) breaking effects.

The pattern of the spectrum in the axial-vector sector~\cite{Guo:2006rp,Albaladejo:2016lbb,Du:2017zvv,Guo:2018gyd} is analogous to the scalar sector. Instead of a single $D_1(2430)$ as listed in the RPP~\cite{ParticleDataGroup:2024cfk}, there exist two poles, one lighter and one heavier. As a consequence of heavy quark spin symmetry, their poles can be obtained using the low-energy constants as those in the scalar sector. The results are $\left(2247_{-6}^{+5}-i107_{-10}^{+11}\right)\,\text{MeV}$ and $\left(2555_{-30}^{+47}-i203_{-9}^{+8}\right)\,\text{MeV}$~\cite{Albaladejo:2016lbb,Du:2017zvv}. Again, they are situated on different sides of the $D_1(2430)$ mass listed in the RPP~\cite{ParticleDataGroup:2024cfk}.
Once again, the lower pole as the SU(3) partner of the $D_{s1}(2460)$ is a member of SU(3) anti-triplet, and the heavier one is rooted in the SU(3) sextet.
In line with the UChPT predictions, LQCD calculations by the Hadron Spectrum Collaboration~\cite{Lang:2022elg,Lang:2025pjq} also find that the ground state axial-vector charmed meson has a relatively small mass. In the coupled-channel LQCD calculation at a pion mass of 391~MeV in Ref.~\cite{Lang:2025pjq}, the lowest $1^+$ charmed meson was found to have a mass of $(2395.6 \pm1.4)\,\text{MeV}$, slightly below the $D^*\pi$ threshold, and smaller than 2.43~GeV. In addition, a heavier pole at $(2737 \pm 79)\,\text{MeV}$ with a substantial width was also found, which could correspond to the heavier $D_1$ pole predicted in UChPT.\footnote{\blue{Recently, the ALICE Collaboration reported their measurements of the $D\pi$ and $D^*(\pi)$ scattering lengths using the femtoscopy technique~\cite{ALICE:2024bhk}.
The isospin-1/2 $D\pi$ and $D^*(\pi)$ scattering lengths are $0.02(1)\,\text{fm}$ and $-0.03(5)\,\text{fm}$, respectively, much smaller than the values extracted from UChPT and lattice QCD results~\cite{Guo:2009ct,Liu:2012zya,Mohler:2012na,Moir:2016srx,Guo:2018kno,Guo:2018tjx,Torres-Rincon:2023qll,Yan:2024yuq}. It is worth mentioning that the two scattering lengths are expected to approximately equal due to heavy quark spin symmetry and the leading order chiral symmetry prediction for these scattering lengths is $\mu/(4\pi F_\pi^2)\approx 0.24$~fm~\cite{Guo:2009ct}, where $\mu$ is the $D^{(*)}\pi$ reduced mass and $F_\pi$ the pion mass decay constant. It gets even enhanced due to the presence of the isospin-1/2 $D_0^*$ ($D_1$) pole pair. Therefore, the small scattering lengths found in the femtoscopy analysis are incompatible with our current understanding of low energy QCD. Alternative experimental determinations of the scattering lengths are urgently called for. }}

Recently, it was argued that in the diquark--anti-diquark tetraquark picture there should also only be flavor [$\bar{3}$] and [6] states, but no [$\overline{15}]$~\cite{Maiani:2024quj}. However, this is correct when only the spin-0 light diquarks alone are considered. When the spin-1 light diquark is also included, which must exist for a consistent diquark
phenomenology~\cite{Barabanov:2020jvn}, both scalar and axial-vector tetraquarks in the $[\overline{15}]$ representation emerge~\cite{Guo:2025imr}, in sharp contrast to the hadronic molecular picture from within the UChPT framework. 
Moreover, the mass difference between scalar and axial-vector diquarks for both heavy-light and light diquarks is similar. As a result of this, in the scalar sector the non-strange members of the
$[\overline{15}]$ multiplet are expected to appear as resonances significantly heavier than the states in the $[6]$. However, in the axial-vector sector, where the lightest states can emerge either from spin-1 and spin-0 diquarks in the heavy-light and light sectors, respectively, or from the spins interchanged, both the non-strange state in the $[\bar 3]$ and in the $[\overline{15}]$ should appear close together. 
One reason for the difference is that the two light antiquarks in the same anti-diquark need to obey the Fermi-Dirac statistics, while such a constraint is absent in the molecular picture since these two light antiquarks are in different mesons. 
Therefore, it is important to compare the flavor structure of the lightest $0^+$ and $1^+$ charmed mesons in an SU(3) symmetric setting. In particular, if e.g. the lattice energy levels appear very similar for the two systems, strong support is provided for a molecular structure of the lowest-lying 
positive-parity open-charm states. On the other hand,
if the two systems
appear to be different, 
strong support is provided for a tetraquark structure---such a result would be inconsistent with a molecular 
structure of the states.

In this paper we present the results of our lattice investigations employing an SU(3) flavor symmetric setting with the pion mass about 613~MeV.
We directly compare the energy levels for the lowest-lying $[6]$ and $[\overline{15}]$ states, finding a high similarity of the two in strong conflict with a tetraquark structure but fully in line with the UChPT predictions for the hadronic molecular structure.

\section{Lattice Calculation}
\subsection{Simulation details}
\label{sec:sim_details}

Our initial studies of the scalar $D\pi$ system aimed at establishing the level structure of the [6] and [$\overline{15}$] states~\cite{Gregory:2021tjs,Gregory:2021rgy}.  By analyzing the energy shifts relative to threshold, we found the [6] to be attractive and [$\overline{15}$] repulsive, consistent with the molecular picture of these states.  Although our analysis could not differentiate between a virtual or bound state in the [6], a later study~\cite{Yeo:2024chk} confirmed our results and established the [6] as a virtual state, while also showing the [$\bar{3}$] to be a deep bound state. 

As argued above,
to decide on the structure of the
positive-parity open-charm states,
it suffices to demonstrate that the level structure of the axial-vector $D^\ast\pi$ system for the [6] and [$\overline{15}$] is the \emph{same} as its scalar analog; that is, the [6] should exhibit attraction and the [$\overline{15}$] repulsion.  In this work we do exactly this.  A more sophisticated L\"uscher analysis will be required to ascertain the exact nature of the [6] (e.g. as was done in~\cite{Yeo:2024chk}) and we leave such work for a later time.

We performed a lattice simulation with $N_f=3+1$ flavors of dynamical clover-improved Wilson fermions~\cite{Sheikholeslami:1985ij} with six iterations of stout smearing~\cite{Morningstar:2003gk}. For both the generation of lattice gauge configurations and correlator measurement, we use the Chroma library for LQCD~\cite{Edwards:2004sx}, with either the QPHIX~\cite{QPHIX,QPHIX2} or QUDA~\cite{Clark:2009wm} linear solver library. We used both GPU-based and CPU-based architectures for this work.

The parameter space target is one where the charm quark has roughly physical mass, and 
the three degenerate light flavors produce a $\sim 600{-}700$~MeV pion, as it is in this range that UChPT  predicts an attractive state in the $[6]$ representation \cite{Du:2017zvv}.

In this SU(3) symmetric setting, there is no mixing among different SU(3) irreps, and thus the axial-vector $D_1(2420)$, which lives in the flavor anti-triplet and couples to $D^*\pi$ in $D$ wave, does not have any influence on our following results on the $[6]$ and [$\overline{15}$] irresps.

This setup is far from the physical point, where LQCD parameters are tuned directly or indirectly with light hadron masses.
We therefore relied on novel methods to reach the target point and determine the lattice spacing. This process required generating a number of short, thermalized tuning ensembles. In this paper we focus on simulations with the lattice coupling constant $\beta=3.6$, which fixes the lattice spacing $a$.
Note that our goal here is only a comparison between the lattice levels in the $0^+$ and $1^+$ sectors. Thus, no continuum extrapolation is needed to extract the desired information.

\begin{figure}
    \centering
    \includegraphics[width=0.49\linewidth]{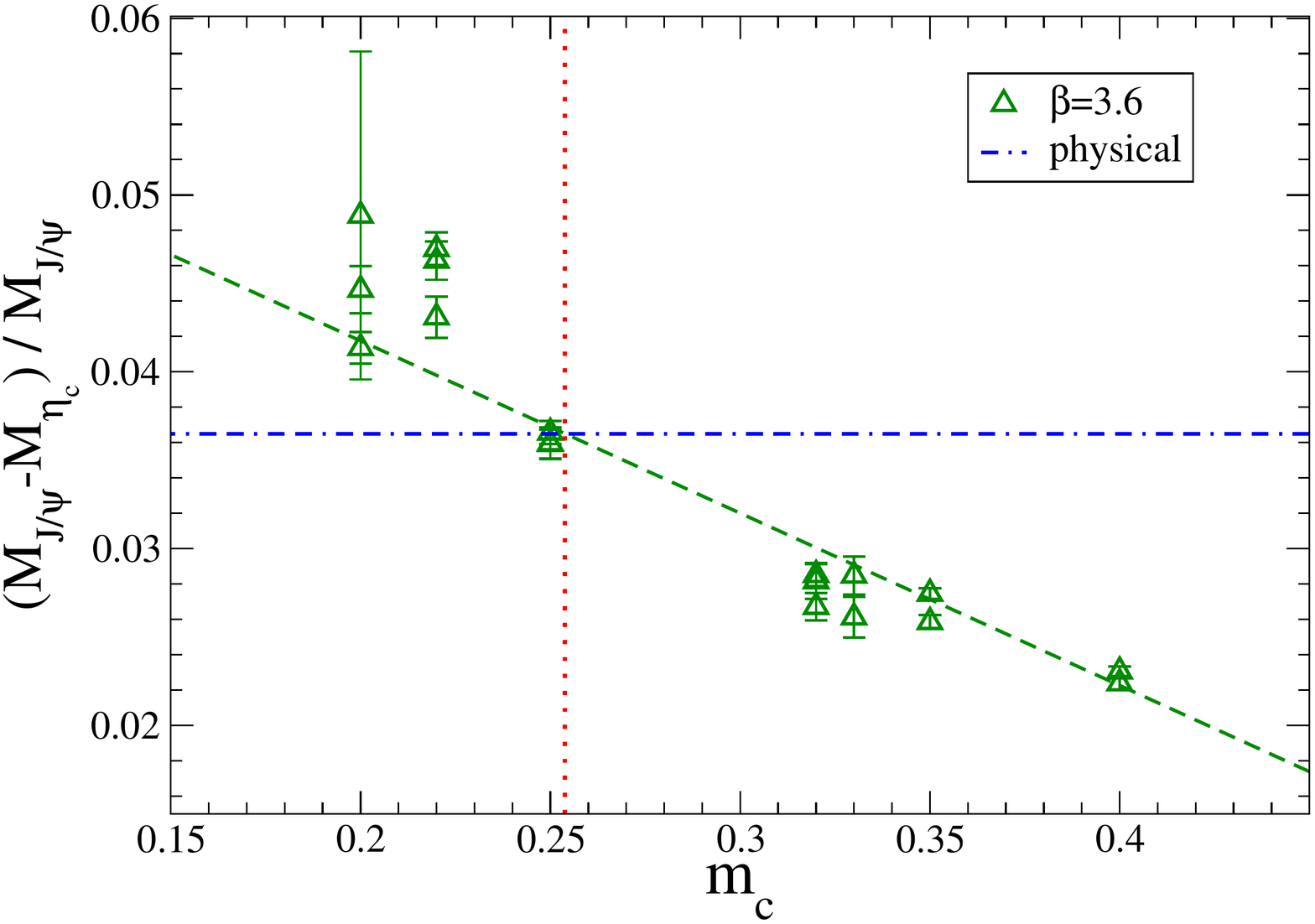}
        \includegraphics[width=0.49\linewidth]{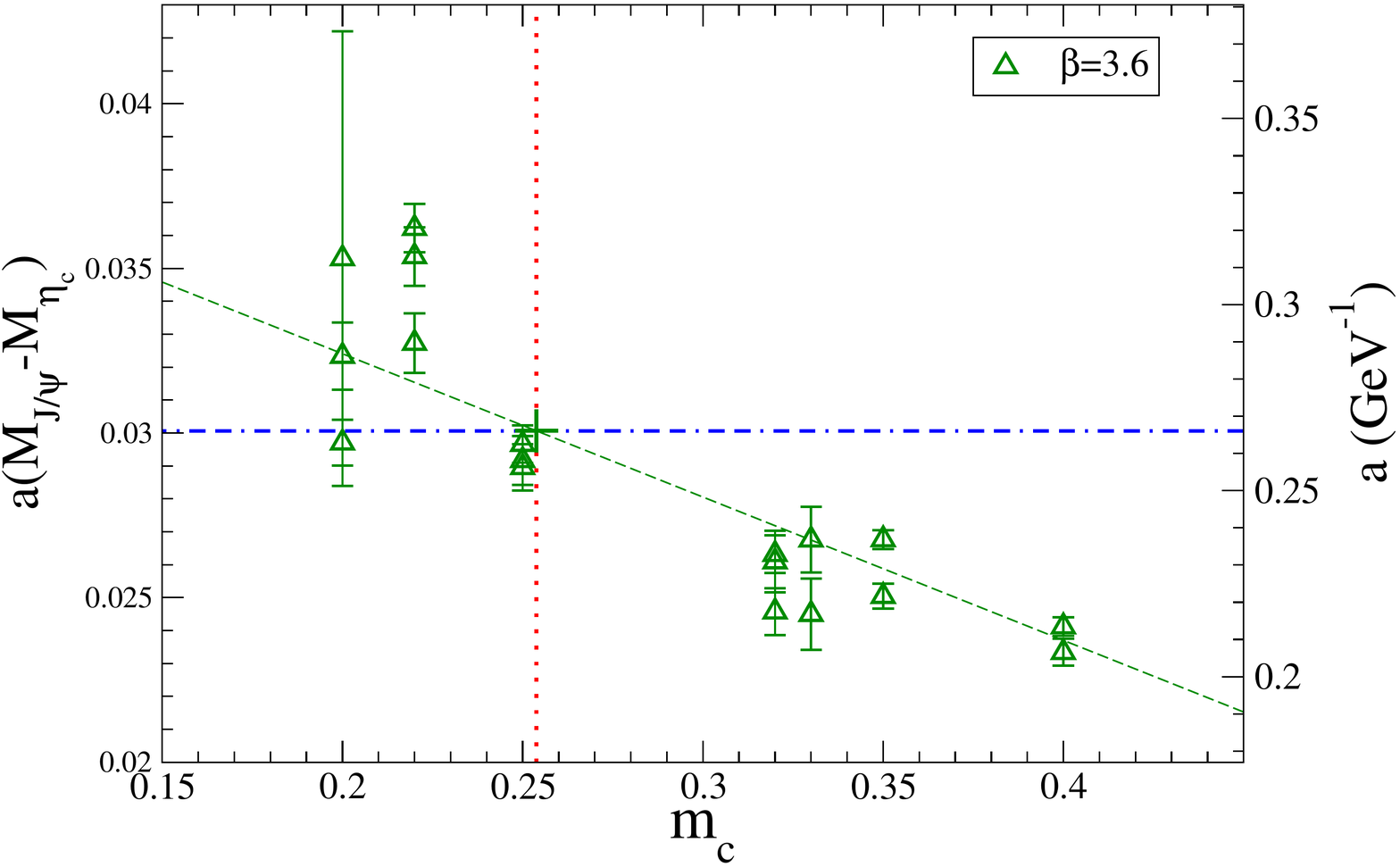}
    \caption{Tuning $m_c$ (left) by varying $m_c$ until $(M_{J/\psi} - M_{\eta_c})/M_{J/\psi}=0.0365$, and establishing the lattice scale $a$ (right) by using the splitting $M_{J/\psi} - M_{\eta_c}$ as a reference. 
    }
    \label{fig:mc_tuning}
\end{figure}

We first vary the charm mass parameter $m_c$ until the dimensionless ratio
 $(M_{J/\psi} - M_{\eta_c})/M_{J/\psi}$ reaches its physical value of $0.0365$.  At this point we can make an estimate of the lattice spacing $a$ with the relation
 \begin{equation}
  \frac{  \left( M_{J/\psi}^{\rm latt} - M_{\eta_c}^{\rm latt}\right)}{\left(M_{J/\psi}^{\rm exp} - M_{\eta_c}^{\rm exp}\right)} =a,
 \end{equation}
 where dimensionless lattice mass parameters are related to physical mass through the lattice spacing $a$ through $aM^{\rm latt} = M^{\rm phys}$, and as a physical value of the splitting we use
 the experimental value $M_{J/\psi}^{\rm exp} - M_{\eta_c}^{\rm exp}=0.113$~GeV.
This fixes $a\approx 0.27$ GeV$^{-1} =$ 0.053 fm for $\beta=3.6$. This process is illustrated in 
\autoref{fig:mc_tuning}. We neglect the effect of disconnected diagrams during the $m_c$ tuning and $a$ determination. Since Wick contractions require the color singlet $c\bar c$ pair to be annihilated and created, thus is highly suppressed by the Okubo-Zweig-Iizuka rule. We estimate that this introduces a maximal systematic uncertainty about 10\% \cite{Levkova:2010ft,Hatton:2020qhk}, which will not affect our conclusions.

\begin{figure}
    \centering
    \includegraphics[width=0.7\linewidth]{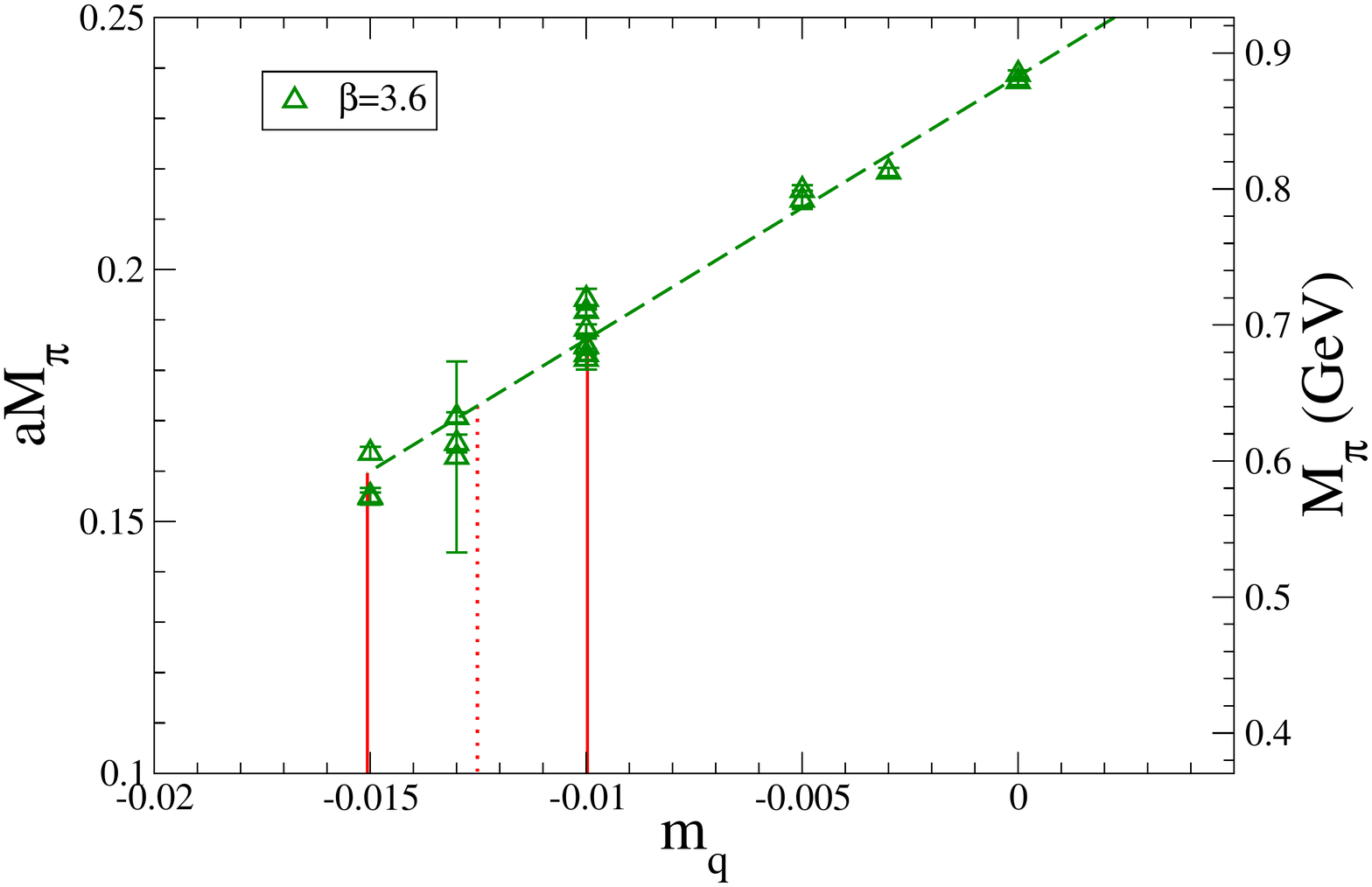}
    \caption{Tuning the light quark mass parameter $m_q$. The solid vertical lines intersect the dashed fit line at the points corresponding to 600 and 700 MeV on the right-hand axis. The vertical dotted line marks the 650 MeV target. We chose an adjacent $m_q=-0.013$ ensemble.}
    \label{fig:mq_tuning}
\end{figure}

Finally, with $m_c$ fixed, we vary the light quark mass parameter $m_q$ until the pion mass falls in the desired range as shown in \autoref{fig:mq_tuning}. We find a target point  ($\beta=3.6$, $m_q=-0.013$, $m_c=0.25$) with
\begin{align*}
        M_\pi&=(613 \pm 1) \text{ MeV},\\
        M_D&=(1890 \pm 2)  \text{ MeV},\\
        M_{D^*}&= (2037 \pm 4)  \text{ MeV}.
\end{align*}
The uncertainties quoted here, and throughout the remainder of this paper are \textit{statistical} only.

At our target parameter values 
we used a hybrid Monte Carlo (HMC) algorithm to  generate an ensemble of lattice volume $64^3\times 64$ with 8400 update trajectories, including 400 thermalization updates. 

\subsection{Interpolating operators and Wick contractions}

The $c$ quark is in an SU(3) singlet, and the remaining degenerate light quarks are projected into either the [6] or $[\overline{15}]$ irrep~\cite{Georgi:1982jb}.  We provide the states and their associated quantum numbers for the [6] and $[\overline{15}]$ irreps in \autoref{tab:6} and \autoref{tab:15}, respectively.
\begin{table}
\center
\caption{The light quark content of the states in the $[6]$ irrep and their associated quantum numbers.  $T^2_a$ is the Casimir operator, $I_z$ is the third component of isospin, and $Y$ is the hypercharge.\label{tab:6}}
{\scriptsize
\begin{tabular}{c|c|c|c|c}
    \hline\hline
state&components& $T_a^2$ & $I_z$ & $Y$\\
\hline\hline
1& $-|u\bar{u}\bar{d}\rangle\frac{1}{2}+|u\bar{d}\bar{u}\rangle\frac{1}{2}-|s\bar{d}\bar{s}\rangle\frac{1}{2}+|s\bar{s}\bar{d}\rangle\frac{1}{2}$ &$\frac{10}{3}$ &$+\frac{1}{2}$ & $-\frac{1}{3}$\\
2& $|d\bar{u}\bar{d}\rangle\frac{1}{2}-|d\bar{d}\bar{u}\rangle\frac{1}{2}-|s\bar{u}\bar{s}\rangle\frac{1}{2}+|s\bar{s}\bar{u}\rangle\frac{1}{2}$ &$\frac{10}{3}$ &$-\frac{1}{2}$ & $-\frac{1}{3}$\\
\hline
3& $|u\bar{u}\bar{s}\rangle\frac{1}{2}-|u\bar{s}\bar{u}\rangle\frac{1}{2}-|d\bar{d}\bar{s}\rangle\frac{1}{2}+|d\bar{s}\bar{d}\rangle\frac{1}{2}$ &$\frac{10}{3}$ &0 & $+\frac{2}{3}$\\
4& $|d\bar{s}\bar{u}\rangle\frac{1}{\sqrt{2}}-|d\bar{u}\bar{s}\rangle\frac{1}{\sqrt{2}}$ &$\frac{10}{3}$ &$-1$ & $+\frac{2}{3}$\\
5& $|u\bar{s}\bar{d}\rangle\frac{1}{\sqrt{2}}-|u\bar{d}\bar{s}\rangle\frac{1}{\sqrt{2}}$ &$\frac{10}{3}$ &$+1$ & $+\frac{2}{3}$\\
\hline
6& $|s\bar{d}\bar{u}\rangle\frac{1}{\sqrt{2}}-|s\bar{u}\bar{d}\rangle\frac{1}{\sqrt{2}}$ &$\frac{10}{3}$ &0 & $-\frac{4}{3}$\\
\hline\hline
\end{tabular}
}
\end{table}

\begin{table}
\center
\caption{The states in the $[\overline{15}]$ irrep and their associated quantum numbers. \label{tab:15}}
\center
{\scriptsize
\begin{tabular}{c|c|c|c|c}
    \hline\hline
    state&components& $T_a^2$ & $I_z$ & $Y$\\
\hline\hline
1& $|s \bar{s}\bar{s}\rangle\frac{1}{\sqrt{3}}-|u \bar{u}\bar{s}\rangle\frac{1}{\sqrt{3}}-|u \bar{s}\bar{u}\rangle\frac{1}{\sqrt{3}}$ &$\frac{16}{3}$ &$0$ & $+\frac{2}{3}$\\
\hline
2& $-|d \bar{u}\bar{d}\rangle\frac{1}{2}-|d \bar{d}\bar{u}\rangle\frac{1}{2}+|s \bar{d}\bar{s}\rangle\frac{1}{2}+|s \bar{s}\bar{d}\rangle\frac{1}{2}$ &$\frac{16}{3}$ &$+\frac{1}{2}$ & $-\frac{1}{3}$\\
3& $-|u \bar{u}\bar{u}\rangle\frac{1}{\sqrt{3}}+|s \bar{u}\bar{s}\rangle\frac{1}{\sqrt{3}}+|s \bar{s}\bar{u}\rangle\frac{1}{\sqrt{3}}$ &$\frac{16}{3}$ &$-\frac{1}{2}$ & $-\frac{1}{3}$\\
\hline
4& $|d \bar{s}\bar{s}\rangle$ &$\frac{16}{3}$ &$-\frac{1}{2}$ & $+\frac{5}{3}$\\
5& $|u \bar{s}\bar{s}\rangle$ &$\frac{16}{3}$ &$+\frac{1}{2}$ & $+\frac{5}{3}$\\
\hline
6& $|d \bar{u}\bar{s}\rangle\frac{1}{\sqrt{2}}+|d \bar{s}\bar{u}\rangle\frac{1}{\sqrt{2}}$ &$\frac{16}{3}$ & $-1$ & $+\frac{2}{3}$\\
7& $|d \bar{d}\bar{s}\rangle\frac{1}{2}+|d \bar{s}\bar{d}\rangle\frac{1}{2}-|u \bar{u}\bar{s}\rangle\frac{1}{2}-|u \bar{s}\bar{u}\rangle\frac{1}{2}$ &$\frac{16}{3}$ & $0$ & $+\frac{2}{3}$\\
8&$|u \bar{d}\bar{s}\rangle\frac{1}{\sqrt{2}}+|u \bar{s}\bar{d}\rangle\frac{1}{\sqrt{2}}$ &$\frac{16}{3}$ & $-1$ & $+\frac{2}{3}$\\
\hline
9& $|s \bar{u}\bar{u}\rangle$ &$\frac{16}{3}$ &$-1$ & $-\frac{4}{3}$\\
10& $|s \bar{u}\bar{d}\rangle\frac{1}{\sqrt{2}}+|s \bar{d}\bar{u}\rangle\frac{1}{\sqrt{2}}$ &$\frac{16}{3}$ &$0$ & $-\frac{4}{3}$\\
11& $|s \bar{d}\bar{d}\rangle$ &$\frac{16}{3}$ &$+1$ & $-\frac{4}{3}$\\
\hline
12& $|d \bar{u}\bar{u}\rangle$ &$\frac{16}{3}$ &$-\frac{3}{2}$ & $-\frac{1}{3}$\\
13& $-|u \bar{u}\bar{u}\rangle\frac{1}{\sqrt{3}}+|d \bar{u}\bar{d}\rangle\frac{1}{\sqrt{3}}+|d \bar{d}\bar{u}\rangle\frac{1}{\sqrt{3}}$ &$\frac{16}{3}$ &$-\frac{1}{2}$ & $-\frac{1}{3}$\\
14& $-|u \bar{u}\bar{d}\rangle\frac{1}{\sqrt{3}}-|u \bar{d}\bar{u}\rangle\frac{1}{\sqrt{3}}+|d \bar{d}\bar{d}\rangle\frac{1}{\sqrt{3}}$ &$\frac{16}{3}$ &$+\frac{1}{2}$ & $-\frac{1}{3}$\\
15& $|u \bar{d}\bar{d}\rangle$ &$\frac{16}{3}$ &$+\frac{3}{2}$ & $-\frac{1}{3}$\\
\hline\hline
\end{tabular}
}
\end{table}
To construct interpolating operators $O_{[d]}$ of either $d=6, \overline{15}$ irrep we need to couple these flavor states to spinors.  We do this by taking any state in either \autoref{tab:6} and \autoref{tab:15} and couple it with the singlet $c$ state, inserting the desired $\Gamma$ matrices at appropriate places to construct the relevant fermion bilinears.  As a simple example, the interpolating operator for the $i=9^\text{th}$ state of the $[\overline{15}]$ irrep is
\begin{displaymath}
    O^{i=9}_{[d=\overline{15}]}(x';x) = [\bar{u}(x')\Gamma_B c(x')][s(x)\Gamma_A \bar{u}(x)]\ ,
\end{displaymath}
where $x'$ and $x$ represent two different spacetime points and $\Gamma_{A,B}$ are in general different.  

Our correlators are then constructed by Wick contraction of $\langle O(y';y)\bar O(x;x)\rangle$ where we assume all sources are located at the same point $x$.  In principle, choosing appropriate different non-local source locations, for example via distillation, would provide a variational calculation of states, ensuring a better signal in our analysis, but for the goals of this paper such ``point-to-all'' correlators suffices.  Future work will include a full variational calculation of these correlators.

We note that the contractions are diagonal within each irrep, i.e. $\langle O^i_{[d]}\bar O^{i'}_{[d]}\rangle\propto \delta_{i,i'}$.  Furthermore, because of the exact SU(3) symmetry in the given setting, the resulting contractions are identical within each [6] and $[\overline{15}]$ irrep.  When comparing the contractions of the [6] and $[\overline{15}]$, we find that they differ by a relative sign in their exchange term,
\begin{multline}\label{eqn:6 contraction}
\langle O^{i}_{[d]}(y';y)\bar{O}^{i}_{[d]}(x;x)\rangle = \\
\text{Tr}\left[\Gamma_A \gamma_5\mathcal{S}^\dag_{y';x}\gamma_5\Gamma_A\mathcal{S}^{}_{y';x}\right]
\text{Tr}\left[\Gamma_B \gamma_5\mathcal{S}^\dag_{y;x}\gamma_5\Gamma_B\mathcal{C}^{}_{y;x}\right]\\
\pm\text{Tr}\left[\Gamma_B \gamma_5\mathcal{S}^\dag_{y;x}\gamma_5\Gamma_A\mathcal{S}^{}_{y';x}\Gamma_A \gamma_5\mathcal{S}^\dag_{y';x}\gamma_5\Gamma_B\mathcal{C}^{}_{y;x}\right]\ .
\end{multline}
Here $\text{Tr}$ refers to the trace over color and spin degrees of freedom and the minus (plus) sign is for the $d=\overline{15}\ (6)$ irrep. $\mathcal{C}$ refers to the charm quark propagator and $\mathcal{S}$ the degenerate light quark propagator. Finally, for the scalar case we set $\Gamma_A=\Gamma_B=\gamma_5$, whereas for the axial vector case we have $\Gamma_A=\gamma_5$ and we average over $\Gamma_B= \gamma_x$, $\gamma_y$, $\gamma_z$.

\subsection{Correlator measurement and analysis}
\label{subsec:measurement}

\begin{figure}
    \centering
    \includegraphics[width=\linewidth]{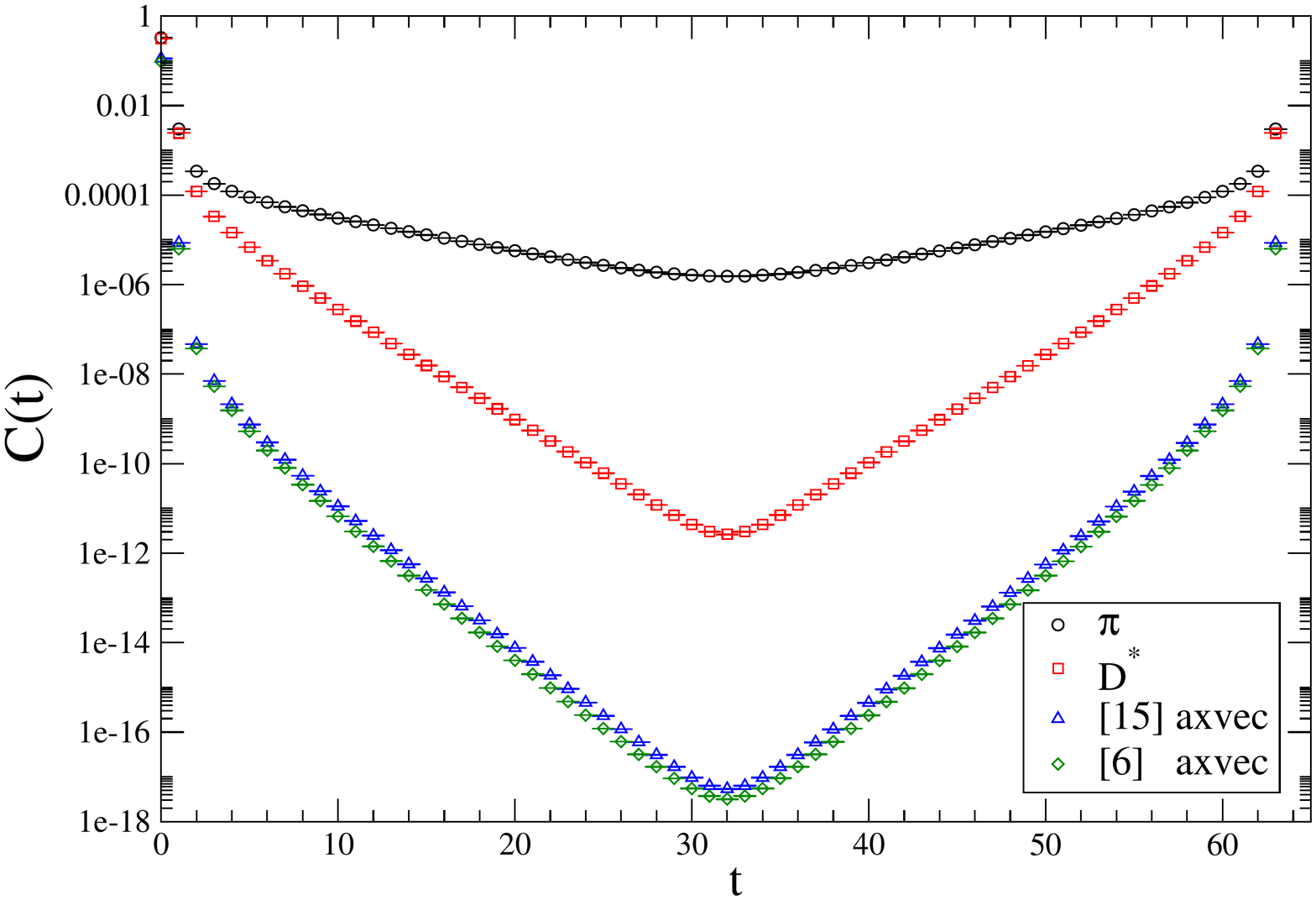}
    \caption{Smeared-sink correlators for the pion, $D^*$ and axial-vector $[\overline{15}]$ and $[6]$ states. The corresponding scalar correlators are qualitatively similar.}
    \label{fig:correlators-example}
\end{figure}

Using the lattice interpolating operators described above,  we measure correlators 
on ensemble described in \autoref{sec:sim_details}. We perform correlator calculations every 40$^\text{th}$ configuration after the thermalization updates. For each measurement we calculate hadron correlators for the $\pi$, $D$, $D^*$, scalar $[6]$ and $[\overline{15}]$ states, and the axial-vector $[6]$ and $[\overline{15}]$ states. On every measurement configuration we calculate correlators  from $N_{\rm src}=128$ different smeared shell sources scattered around the volume with a 4-D Sobol sequence. The correlators are contracted with both point and smeared sinks, giving us two correlators for each state. Example correlators are shown in \autoref{fig:correlators-example}.

To analyze the correlators and get the ground state energies for each state, we fit them to a sum of $N$ exponentials.
For the $q\overline{q}$ states we use
\begin{equation}
    C_{P,s}(t)= \sum^{(N-1)}_{j=0} A_{P,s,j}\cosh\left(E_{P,j}(t-L_t/2)\right),
\end{equation}
where $P={\pi,D,D^*}$. For each state, we simultaneously fit the correlators for both sinks $s\in\{\text{point},\text{smeared}\}$ to extract a common set of energy levels $E_{\pi,j}$, $E_{D,j}$, and $E_{D^*,j}$.
For the positive-parity open-charm states we use:
\begin{multline}\label{eq:fitform}
    C_{P,s}(t) =
    B_{s}\cosh\left((M_{\overline{q}c}-M_\pi)(t-L_t/2)\right)\\
    + \sum_{j=0}^{(N-1)} A_{P,s,j}\cosh\left(E_{p,j}(t-L_t/2)\right).
\end{multline}
Here,  $M_{\overline{q}c}$ and $M_\pi$ are not fit parameters, rather, they come from the meson fits, where $M_\pi$ is the extracted $E_0$ from the pion correlator, and $M_{\overline{q}c}$ is the ground state from the $D$ correlators in the scalar case, and from $D^*$ in the axial-vector case. The $B$ term is a suppressed, but significant, lattice artifact contribution to the correlator. It arises from the $\pi$ propagating forward in time and the $M_{\overline{q}c}$ propagating backwards, or vice-versa.
We fit to $N=2,3,4$ states and vary the fit window $t_{\rm min}\leq t \leq L_t-t_{\rm min}$.  

To achieve a best estimate for the ground state energies of the four states of interest here (scalar and axial-vector $D$ mesons either in the $[6]$ or in the $[\overline{15}]$ irrep) we perform a curated model-averaging over all relevant 2-, 3- and 4-state fits.
The extracted $E_0$ values 
 were plotted as a function of $t_{\rm min}$. For each $N$ and each state, we identify a value  $t_{\rm min}^{\rm cut}$, below which $E_0$ has not yet reached a plateau. For $t_{\rm min}<t_{\rm min}^{\rm cut}$, we discard the fits. The rising $E_0$ values at low $t_{\rm min}$ are an unmistakable signal of contamination by excited states which are not well modeled by the $N$ exponentials in the fit anzatz. The $E_0$ values from the remaining fits are combined in a weighted average using the Akaike Information Criterion (AIC)~\cite{Jay:2020jkz,Neil:2022joj}, which favors lower $\chi^2$ values, fewer fit parameters, and fits that include more of the data.

 \begin{figure*}[tb]
    \centering
     \includegraphics[width=0.49\textwidth]{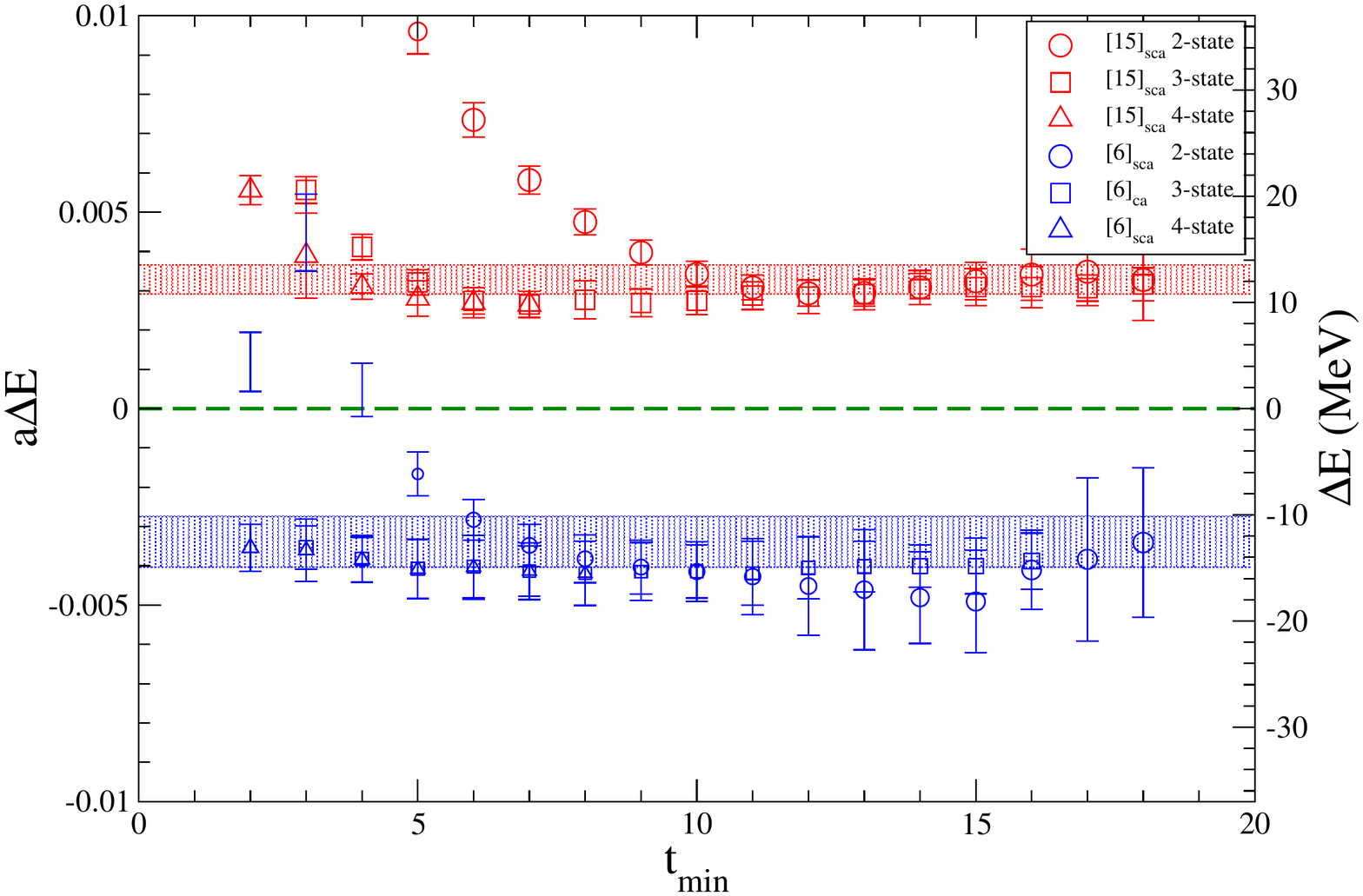}
     \includegraphics[width=0.49\textwidth]{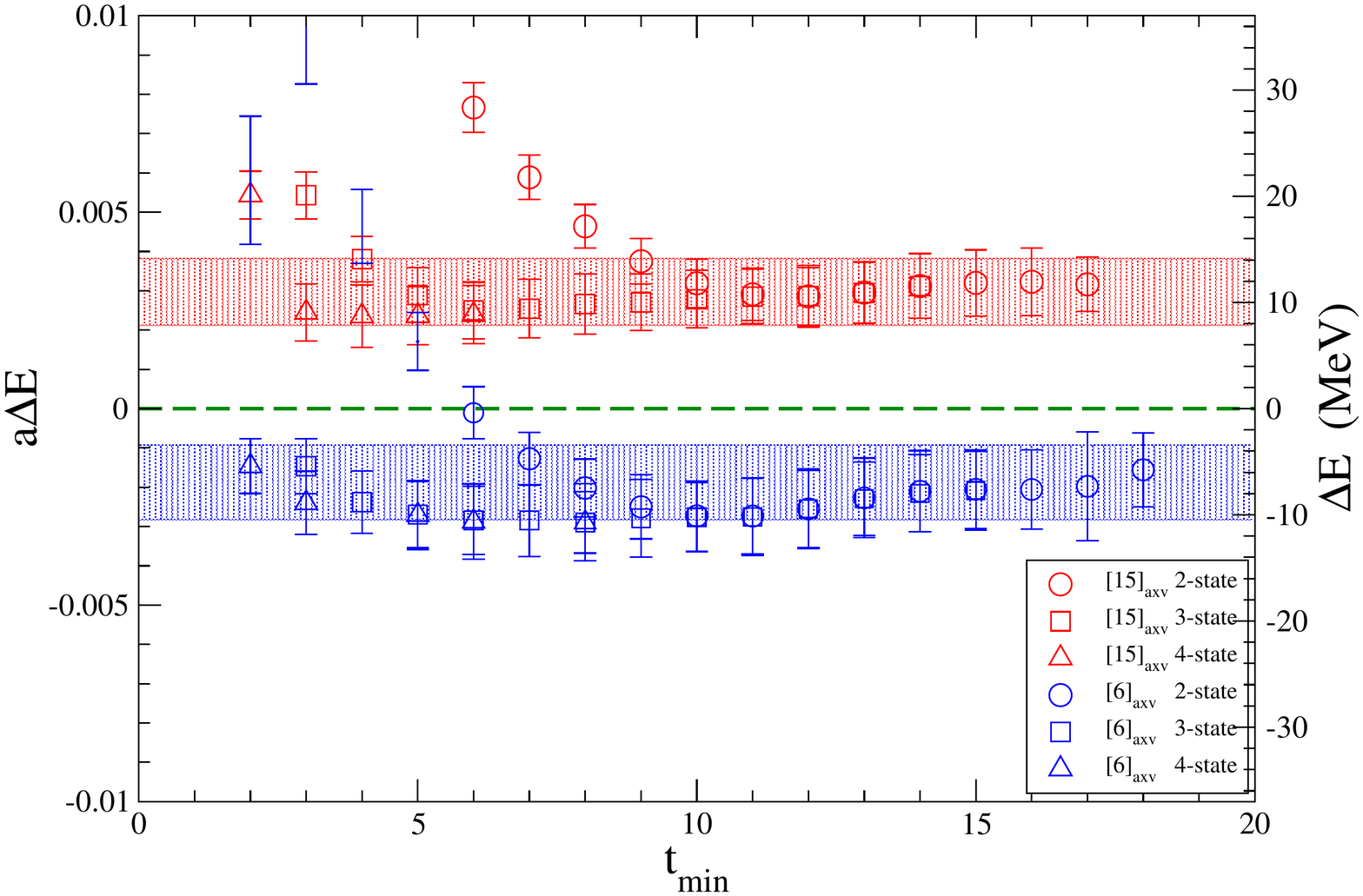}
       \caption{Values of $\Delta E$ as a function of $t_{\rm min}$ fit window limit, from fits to the SU(3) scalar (left panel) and axial-vector (right panel) $[\overline{15}]$  (red), and $[6]$ (blue) representation correlators, using $N=2,3,4$ exponential states (circles, squares, and triangles, respectively). 
       For all of the displayed fits the symbol size is roughly proportional to the fit's $p-$value from a naive $\chi^2$ after convergence. Poor fits have a small or invisible symbol.
       The shaded bands represent the 1 standard deviation interval of the model-averaged $\Delta E$ values.}
     \label{fig:scalar-fits}
     
 \end{figure*}  
   
Finally, we produce binned jackknife samples of the correlator data and repeat the entire fitting procedure and model averaging
on each sample to produce statistical uncertainty estimates on each of the ground-state energies and the energy shifts $\Delta E = E_0- (M_{\overline{q}c}+M_\pi)$. 
The $\Delta E$ values from fits of the SU(3) $[\overline{15}]$ and $[6]$ representation states in the scalar and axial-vector sectors are shown in the left and right panels of \autoref{fig:scalar-fits}, respectively. They
show extremely similar behavior: In both sectors there is clear indication of attraction in the $[6]$ irrep with 
\begin{align}
\Delta E_{{[6]}_{\rm sca}}&= (-13\pm 2)\text{ MeV}\ ,\\
    \Delta E_{{[6]}_{\rm axv}}&=(-7\pm 4)\text{ MeV}\ 
\end{align}
for the scalar and axial-vector sectors, respectively.
On the other hand, in both sectors the $[\overline{15}]$ irrep  presents itself as repulsive
with a positive energy shift 

\begin{align}
    \Delta E_{[\overline{15}]_{\rm sca}} &= (12\pm 1)~\text{ MeV}\ , \notag\\
    \Delta E_{[\overline{15}]_{\rm axv}} &= (11\pm 3)~\text{ MeV}\  .
\end{align} 
We reiterate that the stated uncertainties are statistical only.
This near equality of the two sectors is reproducing the predictions from UChPT, which predicts a molecular structure, while is clearly at 
odds with the tetraquark picture as detailed in the introduction.

\section{Conclusions \& Discussion}
\label{sec:conclusions}

In this lattice study we demonstrate that the lowest energy levels for four-quark systems in the SU(3) flavor $[6]$
and $[\overline{15}]$ multiplets
behave almost the same for scalar and axial-vector quantum numbers. As detailed in Ref.~\cite{Guo:2025imr}
this behavior is consistent with expectations within the molecular picture while being at odds with those for a diquark--anti-diquark structure, as soon as both 
scalar and axial-vector light diquarks are included. Thus, the only way to reconcile
our findings with the compact tetraquark picture is to abandon spin-1 light diquarks as relevant degrees of freedom within positive-parity singly-heavy mesons. However, for us this seems to be quite unnatural, for it would question also the diquark phenomenology of singly-heavy baryons and thus as diquarks as relevant degrees of freedom within hadrons. To summarize, our results provide strong support for a molecular structure of the ground state positive-parity open-charm
states and exclude a conventional quark--antiquark structure as well as 
a diquark--anti-diquark structure. 

\section*{Acknowledgements}
We are grateful to Giovanni Pederiva for useful discussion on data modeling.
T.L. and S.K. were funded in part by the Deutsche
	Forschungsgemeinschaft (DFG, German Research Foundation) as part of the CRC 1639 NuMeriQS–project no.~511713970. S.K. is further supported by the MKW NRW under funding code NW21-024-A.
The authors gratefully acknowledge computing time on the supercomputer JURECA\cite{JURECA} at Forschungszentrum Jülich under Grant ``EXFLASH'', and also 
the Gauss Centre for Supercomputing e.V. (\url{www.gauss-centre.eu}) for funding this project by providing computing time through the John von Neumann Institute for Computing (NIC) on the GCS Supercomputer JUWELS\cite{JUWELS} at Jülich Supercomputing Centre (JSC).
E.B.G. was supported in part by the German Federal Ministry of Education and
Research (BMBF) and the Ministry of Culture and Science (MKW) of the
state of North-Rhine-Westphalia through funding of the Gauss Centre for
Supercomputing (GCS).
F.-K.G. was supported by the National Key R\&D Program of China under Grant No. 2023YFA1606703, by the National Natural Science Foundation of China (NSFC) under Grants No. 12125507 and No. 12047503, and by the Chinese Academy of Sciences (CAS) under Grant No.~YSBR-101. In addition, C.H. thanks the CAS President's International Fellowship Initiative (PIFI) under Grant No.~2025PD0087.
\bibliographystyle{epj}
\bibliography{refs}
%

\end{document}